\documentclass[aps, prl, superscriptaddress, showpacs,
footinbib, 
amsmath,
preprintnumbers,
twocolumn,
floatfix]{revtex4-1}

\usepackage{amsmath}
\usepackage{graphicx}

\usepackage{longtable}

\usepackage{bm}
\usepackage[usenames,dvipsnames]{color}
\usepackage{verbatim}
\usepackage{amsfonts}
\usepackage{longtable}
\usepackage{natbib}
\usepackage{hyperref}

\usepackage{refcount}

\usepackage{dcolumn}

\begin{document}
 \title{{\bf 
Novel $|V_{us}|$ Determination Using Inclusive Strange $\tau$ Decay and 
Lattice HVPs}
 \vspace{5mm}}

\newcommand\edinb{SUPA, School of Physics, The University of Edinburgh, 
Edinburgh EH9 3JZ, UK}
\newcommand\soton{School of Physics and Astronomy, University of Southampton, 
Southampton SO17 1BJ, UK}
\newcommand\bnl{Physics Department, Brookhaven National Laboratory, Upton, 
NY 11973, USA}

\author{Peter Boyle}\affiliation{\edinb}
\author{Renwick James Hudspith} 
      \affiliation{York University, 4700 Keele St., Toronto, ON Canada M3J IP3}
\author{{Taku Izubuchi}} 
      \affiliation{\bnl} 
      \affiliation{RIKEN-BNL Research Center, Brookhaven National Laboratory, 
      Upton, NY 11973, USA} 
\author{Andreas J\"uttner}\affiliation{\soton}
\author{Christoph Lehner}\affiliation{\bnl} 
\author{Randy Lewis} 
      \affiliation{York University, 4700 Keele St., Toronto, ON Canada M3J IP3}
\author{Kim Maltman} 
      \affiliation{York University, 4700 Keele St., Toronto, ON Canada M3J IP3}
      \affiliation{CSSM, University of Adelaide, Adelaide, SA 5005 Australia}
\author{Hiroshi Ohki} 
      \affiliation{RIKEN-BNL Research Center, Brookhaven National Laboratory, 
Upton, NY 11973, USA}
      \affiliation{Physics Department, Nara Women's University, Nara 630-8506, Japan}
\author{Antonin Portelli}
\affiliation{School of Physics and Astronomy, University of Edinburgh, 
Edinburgh EH9 3JZ, United Kingdom} 
\author{Matthew Spraggs}
\affiliation{School of Physics and Astronomy, University of Edinburgh, 
Edinburgh EH9 3JZ, United Kingdom} 
\collaboration{ RBC and UKQCD Collaborations}

\begin{abstract}
We propose and apply a new approach to determining $|V_{us}|$ using dispersion
relations with weight functions having poles at Euclidean (space-like) 
momentum which relate strange hadronic $\tau$ decay distributions to 
hadronic vacuum polarization functions (HVPs) obtained from lattice 
QCD. We show examples where spectral integral contributions from the region
where experimental data have large errors or do not exist are strongly
suppressed but accurate determinations of the relevant lattice HVP
combinations remain possible. 
The resulting $|V_{us}|$ agrees well with determinations from $K$ 
physics and 3-family CKM unitarity. Advantages of this new approach 
over the conventional hadronic $\tau$ decay determination 
employing flavor-breaking sum rules are also discussed.
\end{abstract}
\maketitle
\section{ Introduction}
Precise determinations of the Cabibbo-Kobayashi-Maskawa (CKM) matrix 
element $|V_{us}|$ are important in the context of 3-family unitarity 
tests and searches for physics beyond the Standard Model (SM). The 
most precise such determination, $|V_{us}|=0.2253(7)$, is from 
$\Gamma [K_{\mu2}]/\Gamma [\pi_{\mu2}]$, using the lattice input
$f_K/f_\pi=1.193(3)$~\cite{Dowdall:2013rya, Bazavov:2014wgs, Carrasco:2014poa, Aoki:2016frl}. Three-family 
unitarity and $\vert V_{ud}\vert = 0.97417(21)$~\cite{Towner:2014uta},
similarly, imply $|V_{us}|=0.2258(9)$, while $K_{l3}$, with lattice
$f_+(0)$ input, yields $|V_{us}|=0.2237(10)$~\cite{Olive:2016xmw}. 
It is a long-standing puzzle that conventional flavor-breaking (FB) 
finite-energy sum rules (FESRs) employing hadronic $\tau$ decay data 
yield much lower $\vert V_{us}\vert$, most recently 
$0.2186(21)$~\cite{HFLAV-Tau2017} ($0.2207(27)$ when $\Gamma [K_{\mu 2}]$
and dispersive $K\pi$ form factor constraints
are incorporated~\cite{Antonelli:2013usa}).

The conventional FB FESR implementation employs assumptions for unknown 
dimension $D=6$ and $8$ OPE condensates which turn out to fail 
self-consistency tests~\cite{Hudspith:2017vew}. An alternate implementation,
fitting $D>4$ condensates to data, yields results passing these tests and 
compatible with determinations from other sources~\cite{Hudspith:2017vew}. The 
resulting error is dominated by uncertainties on the relevant weighted 
inclusive flavor $us$ spectral integrals and a factor $>2$ larger than 
that of $K$-decay-based approaches. Improved branching fractions (BFs) 
used in normalizing low-multiplicity $us$ exclusive-mode Belle and BaBar 
distributions would help, but $\sim 25\%$ errors on higher-multiplicity $us$
``residual mode'' contributions~\cite{Barate:1999hj}, involving modes not 
re-measured at the B-factories, preclude a factor of $2$ 
improvement~\cite{Hudspith:2017vew,hilmoz16}.

This paper presents a novel dispersive approach to determining 
$\vert V_{us}\vert$ using inclusive strange hadronic $\tau$ decay 
data, hadronic vacuum polarization functions (HVPs) computed on the 
lattice, and weight functions, $\omega_N (s)=\Pi_{k=1}^N(s+Q_k^2)^{-1}$, 
$Q_k^2>0$, having poles at Euclidean $Q^2=Q_k^2>0$. We show examples 
of such $\omega_N$ which strongly suppress spectral contributions from the 
high-multiplicity $us$ ``residual'' region without blowing up errors 
on the related lattice HVP combinations. The approach yields 
$\vert V_{us}\vert$ in good agreement with $K$-decay analysis 
results and 3-family CKM unitarity expectations.
The lattice error is comparable to the experimental one, 
and the total error is less than that of the inclusive FB FESR $\tau$ decay determination.

\section{New inclusive determination}
The conventional inclusive FB $\tau$ decay determination  
is based on the FESR relation~\cite{Braaten:1991qm, LeDiberder:1992zhd}
\begin{align}\label{eq:dr}
\int_0^{s_0} \omega(s) \Delta\rho(s) ds = -\frac{1}{2\pi i} 
\oint_{|s|=s_0} \omega(s) \Delta\Pi(-s) ds, 
\end{align}
connecting, for any $s_0$ and analytic $\omega (s)$, the relevant FB 
combination, $\Delta\Pi (-s)=\Pi_{us}(-s) -\Pi_{ud}(-s)$, 
of spin $J=0,1$ HVPs and associated spectral function 
$\Delta\rho (s)=\frac{1}{\pi}\rm{Im}\Delta\Pi(-s)$.
Experimental data is used on the LHS and, for large enough $s_0$, the OPE 
on the RHS. In the SM, the differential distribution, $dR_{V/A;ij}/ds$, 
associated with the flavor $ij=ud,us$ vector $(V)$ or axial vector $(A)$ 
current-induced decay ratio 
$R_{V/A;ij}=\Gamma[\tau^-\to\nu_\tau \rm{hadrons}_{V/A;ij}]/ 
\Gamma[\tau^- \to e^- \bar{\nu}_e\nu_\tau]$, is related to the $J=0,1$ 
spectral functions $\rho^{(J)}_{ij;V/A}(s)$, by~\cite{tsai}
\begin{align}\label{eq:bf}
\frac{d R_{ij;V/A} }{ds}
=& \frac{12\pi^2 |V_{ij}|^2 S_{EW}}{m_\tau^2}
\\ \nonumber 
& \times \left[
\omega_\tau(s) \rho^{(0+1)}_{ij;V/A}(s)
-\omega_L (y_\tau) \rho^{(0)}_{ij;V/A}(s)
\right], 
\end{align}
where  $y_\tau=s/m_\tau^2$, $\omega_\tau(y)=(1-y)^2 (1+2y)$, 
$\omega_L(y)=2y(1-y)^2$, and $S_{EW}$ is a known short-distance 
electroweak correction~\cite{Marciano:1988vm, Erler:2002mv}. 
Experimental $dR_{ij;V/A}/ds$ distributions thus determine,
up to factors of $\vert V_{ij}\vert^2$, combinations of the 
$\rho^{(J)}_{ij;V/A}$.

The low $\vert V_{us}\vert$ noted above results from a conventional 
implementation~\cite{pichetal} of Eq.~(\ref{eq:dr}) which employs fixed 
$s_0=m_\tau^2$ and $\omega =\omega_\tau$ and assumptions for experimentally 
unknown $D=6$ and $8$ condensates. With $s_0=m_\tau^2$ and 
$\omega =\omega_\tau$, inclusive non-strange and strange BFs determine
the $ud$ and $us$ spectral integrals. Testing $D=6$ and $8$ assumptions 
by varying $s_0$ and/or $\omega$, however, yields $\vert V_{us}\vert$ 
with significant unphysical $s_0$- and $\omega$-dependence, motivating 
an alternate implementation employing variable $s_0$ and $\omega$ which 
allows a simultaneous fit of $\vert V_{us}\vert$ and the $D>4$ condensates. 
Significantly larger (now stable) $\vert V_{us}\vert$ are found, the 
conventional implementation results 
$\vert V_{us}\vert =0.2186(21)$~\cite{HFLAV-Tau2017} and
$0.2207(27)$~\cite{Antonelli:2013usa}, shifting up to $0.2208(23)$
and $0.2231(27)$~\cite{Hudspith:2017vew}, respectively, with the new implementation.
$us$ spectral integral uncertainties dominate the
error, with current $\sim 25\%$ residual mode contribution errors
precluding a competitive determination~\cite{Hudspith:2017vew}.

Motivated by this limitation, we switch to generalized dispersion 
relations involving the experimental $us$ V+A inclusive distribution 
and weights, $\omega_N(s)\equiv \prod_{k=1}^N\left( s+Q_k^2\right)^{-1}$,
$0<Q_k^2<Q_{k+1}^2$, having poles at $s=-Q_k^2$. 
From Eq.~(\ref{eq:bf}), $dR_{us; V+A}/ds$ directly determines
$\vert V_{us}^2\vert\, \tilde{\rho}_{us}(s)$, with 
\begin{eqnarray}
\tilde{\rho}_{us}(s) \equiv \left( 1+2\frac{s}{m_\tau^2} \right) 
\rho^{(1)}_{us; V+A}(s) + \rho^{(0)}_{us; V+A}(s)\ .
\end{eqnarray}
For $N \ge 3$, the associated HVP combination
\begin{eqnarray}\label{eq:lathvp}
\tilde{\Pi}_{us} \equiv
\left(  1-2\frac{Q^2}{m_\tau^2}\right) \Pi_{us;V+A}^{(1)}(Q^2) + 
\Pi_{us;V+A}^{(0)}(Q^2) 
\end{eqnarray}
satisfies the convergent dispersion relation
\begin{eqnarray} \label{eq:gd}
&&\int_0^\infty \tilde{\rho}_{us}(s) \omega_N(s) ds = 
\sum_{k=1}^N \underset{s=-Q_k^2}{\rm Res} 
\left[ \tilde{\Pi}_{us}(-s) 
\omega_N(s)\right] 
\nonumber\\
&&\qquad\qquad =\sum_{k=1}^N {\frac{ \tilde{\Pi}_{us;V+A}(Q_k^2)}{\prod_{j\ne k}
\left( Q_j^2-Q_k^2\right)}} \equiv \tilde{F}_{\omega_N}\, .
\end{eqnarray}
With $\tilde{\Pi}_{us}(Q^2_k)$ measured on the lattice, $dR_{us;V+A}/ds$ 
used to fix $s<m_\tau^2$ spectral integral contributions, and
$s > m_\tau^2$ contributions approximated using pQCD, one has, 
\begin{align}\label{eq:vus}
\vert V_{us}\vert  = \sqrt{\tilde{R}_{us;w_N}/\left( 
\tilde{F}_{\omega_N}
-\int_{m_\tau^2}^\infty \tilde{\rho}_{us}^{\rm{pQCD}}(s) \omega_N(s) ds
\right)}\, .
\end{align}
where $\tilde{R}_{w_N} \equiv {\frac{m_\tau^2}{12\pi^2S_{EW}}} 
\int_0^{m_\tau^2}{\frac{1}{(1-y_\tau )^2}} 
{\frac{dR_{us;V+A}(s)}{ds}} \omega_N(s) ds$.

Choosing uniform pole spacing $\Delta$, $\omega_N$ can be characterized
by $\Delta$, $N$, and the pole-interval midpoint, $C=(Q_1^2+Q_N^2)/2$. 
With large enough $N$, and all $Q_k^2$ below $\sim 1$ GeV$^2$, spectral 
integral contributions from $s>m_\tau^2$ and the higher-$s$, larger-error 
part of the experimental distribution can be strongly suppressed. Increasing 
$N$ lowers the error of the LHS in Eq.~\ref{eq:gd}, but increases the 
relative RHS error. With results insensitive to modest changes of $\Delta$, 
we fix $\Delta = 0.2/(N-1)$ GeV$^2$, ensuring $\omega_N$ with the same 
$C$ but different $N$ have poles spanning the same $Q^2$ range. $C$ and $N$ 
are varied to minimize the error on $|V_{us}|$.

We employ the following $us$ spectral input: $K_{\mu2}$ or
$\tau\rightarrow K\nu_\tau$~\cite{HFLAV-Tau2017} for $K$ pole contributions, 
unit-normalized Belle or BaBar distributions for 
$K\pi$~\cite{Epifanov:2007rf,Aubert:2007jh}, 
$K^- \pi^+\pi^-$~\cite{Nugent:2013ij},
$\bar{K}^0 \pi^-\pi^0$~\cite{Ryu:2013lca} and 
$\bar{K}\bar{K}K$~\cite{Aubert:2007mh,Lee:2010tc}, the most recent HFLAV 
BFs~\cite{HFLAV-Tau2017}, and 1999 ALEPH results~\cite{Barate:1999hj}, 
modified for current BFs, for the residual mode distribution. 
Multiplication of a unit-normalized distribution by the ratio of 
corresponding exclusive-mode to electron BFs, converts that distribution 
to the corresponding contribution to $dR_{us;V+A}(s)/ds$. The dispersively 
constrained $K\pi$ BFs of Ref.~\cite{Antonelli:2013usa} (ACLP) provide an 
alternate $K\pi$ normalization. In what follows, we illustrate the 
lattice approach using the HFLAV $K\pi$ normalization. Alternate results 
using the ACLP normalization are given in Ref.~\cite{footnote1}.

\section{Lattice calculation method}
We compute the two-point functions of the flavor $us$ V and A currents, 
$J^\mu_{V/A}(\vec{x},t)=J^\mu_{V/A}(x)$, $\mu =x, y, z, t$, via
\begin{eqnarray}
C_{V/A}^{\mu\nu}(t) = \sum_{\vec{x}} \langle J^\nu_{V/A} (\vec{x},t) 
(J^\mu_{V/A}(0,0))^\dagger \rangle\, .
\end{eqnarray}
The continuum spin $J=0,\, 1$, $us$ HVPs, 
$\Pi^{(J)}_{us;V/A}\equiv \Pi^{(J)}_{V/A}$, are related to the 
two-point functions by
\begin{eqnarray}
\sum_{x} e^{iqx} \langle J^\mu_{V/A} (x) (J^\nu_{V/A}(0))^\dagger \rangle 
= \quad \quad \quad \quad \quad \\ \nonumber 
(q^2 g^{\mu\nu} - q^\mu q^\nu) \Pi^{(1)}_{V/A}(q^2) + 
q^\mu q^\nu\Pi^{(0)}_{V/A}(q^2)\, ,
\end{eqnarray}
up to finite volume (FV) and discretization corrections.
With $C_{V/A}^{(1)}(t)=\frac{1}{3}\sum_{k=x,y,z} C_{V/A}^{kk}(t)$,
$C_{V/A}^{(0)}(t)=C_{V/A}^{tt}(t)$, the analogous $J=0,1$ parts of 
$C_{V/A}^{\mu\nu}$ are~\cite{Bernecker:2011gh, Feng:2013xsa}
\begin{align}\label{eq:ft}
\Pi_{V/A}^{(J)}(Q^2)-\Pi_{V/A}^{(J)}(0) = \sum_t 
K(q,t)\, C_{V/A}^{(J)}(t)\, ,
\end{align}
with $K(q,t)=\frac{\cos{qt}-1}{\hat{q}^2}+\frac{1}{2}t^2 $ and
$\hat{q}$ the lattice momentum, $\hat{q}_\mu=2 \sin{q_\mu/2}$.
FV corrections to this infinite volume result are discussed below.
We use lattice HVPs measured on
the near-physical-quark-mass, $2+1$ flavor $48^3 \times 96$ and 
$64^3 \times 128$ M\"obius domain wall fermion ensembles of the
RBC and UKQCD collaborations~\cite{Blum:2014tka}, employing
all-mode-averaging (AMA)~\cite{Blum:2012uh, Shintani:2014vja} 
to reduce costs. Slight $u,\, d,\, s$ mass mistunings are
corrected by measuring the HVPs with partially quenched 
(PQ) physical valence quark masses~\cite{Blum:2014tka}, also
using AMA. 
\begin{figure}[t]
\begin{center}
\includegraphics[scale=0.3,clip]{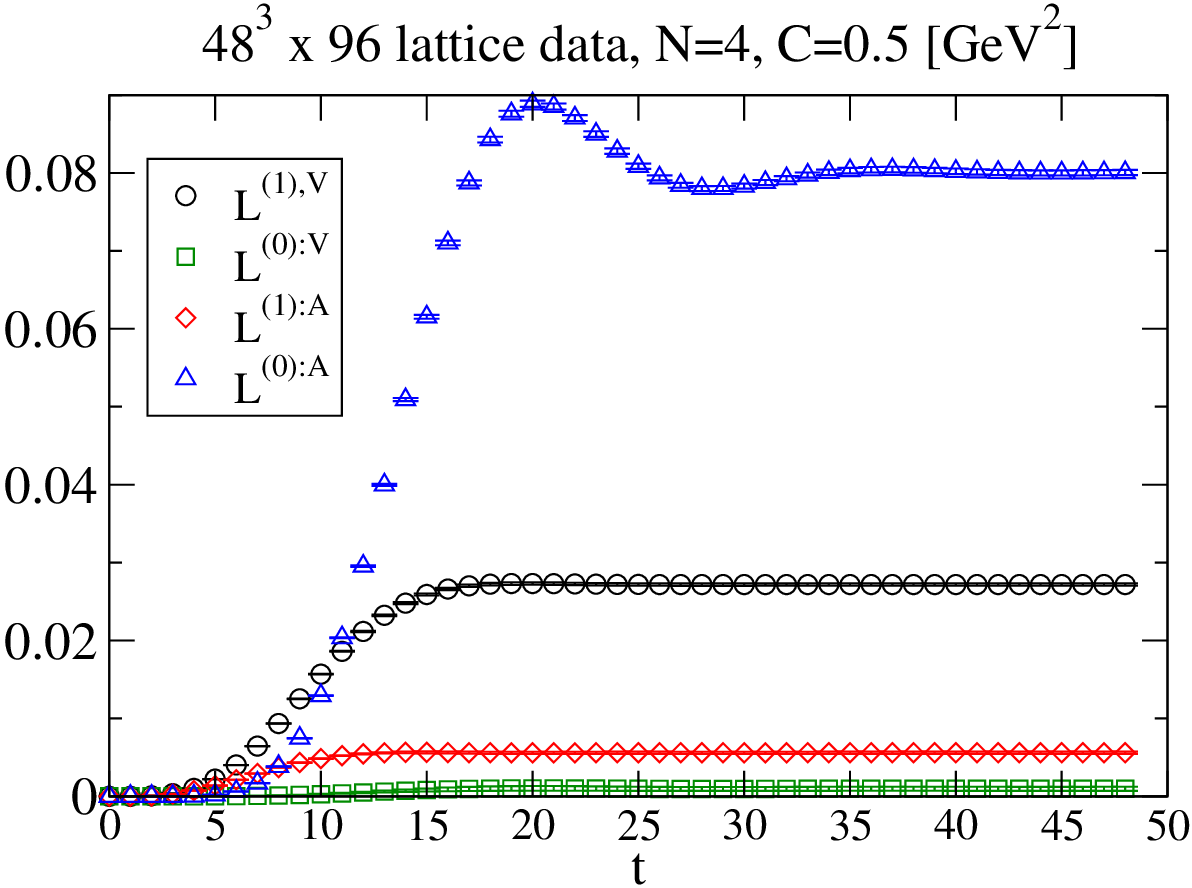} 
\caption{ 
Partial sum of the residues $L_{V/A;\omega_N}^{(J)}(t)$.
}
\label{fig:Lt_48c}
\end{center} 
\end{figure}

$\tilde{F}_{\omega_N}$ in Eq.~(\ref{eq:gd}) can be decomposed into four 
contributions, $\tilde{F}_{V/A; \omega_N}^{(J)}$, labelled by the spin
$J=0$ or $1$, and current type, V or A. 
$\tilde{F}_{V/A; \omega_N}^{(J)}=\lim_{t \to \infty} L_{V/A; 
\omega_N}^{(J)}(t)$,  
where $L_{V/A; \omega_N}^{(J)}(t) = \sum_{l=-t}^{t} \omega^{(J)}_N(l) 
C^{(J)}_{V/A}(l)$. From Eqs.~(\ref{eq:ft}), (\ref{eq:lathvp}), 
\begin{align}
\omega^{(1)}_N  =& 
\sum_{k=1}^N 
K\left( \sqrt{Q_k^2},t\right)
\left( 1-\frac{2 Q^2_k}{m_\tau^2} \right) 
\underset{s=-Q_k^2}{\rm Res} \left[\omega_N(s) \right],
\\ \nonumber
\omega^{(0)}_N  =& 
\sum_{k=1}^{N} 
K\left( \sqrt{Q_k^2},t\right)
\underset{s=-Q_k^2}{\rm Res} \left[\omega_N(s) \right].
\end{align}
With finite lattice temporal extent, finite-time effects may exist.
Increasing $N$ increases the level of cancellation and 
relative weight of large-$t$ contributions on the RHS of Eq.~(\ref{eq:gd}).
The restrictions $0.1\ {\rm GeV}^2<C<1\ {\rm GeV}^2$ and $N \le 5$,
chosen to strongly suppress large-$t$ contributions, allow us to avoid 
modelling the large-$t$ behavior. Fig.~\ref{fig:Lt_48c} shows, as an 
example, the large-$t$ plateaus of the partial sums 
$L_{V/A; \omega}^{(J)}(t)$, obtained in all four channels, for $N=4$, 
$C=0.5$ GeV$^2$ on the $48^3 \times 96$ ensemble. 

\begin{figure}[t]
\includegraphics[scale=0.3,clip]{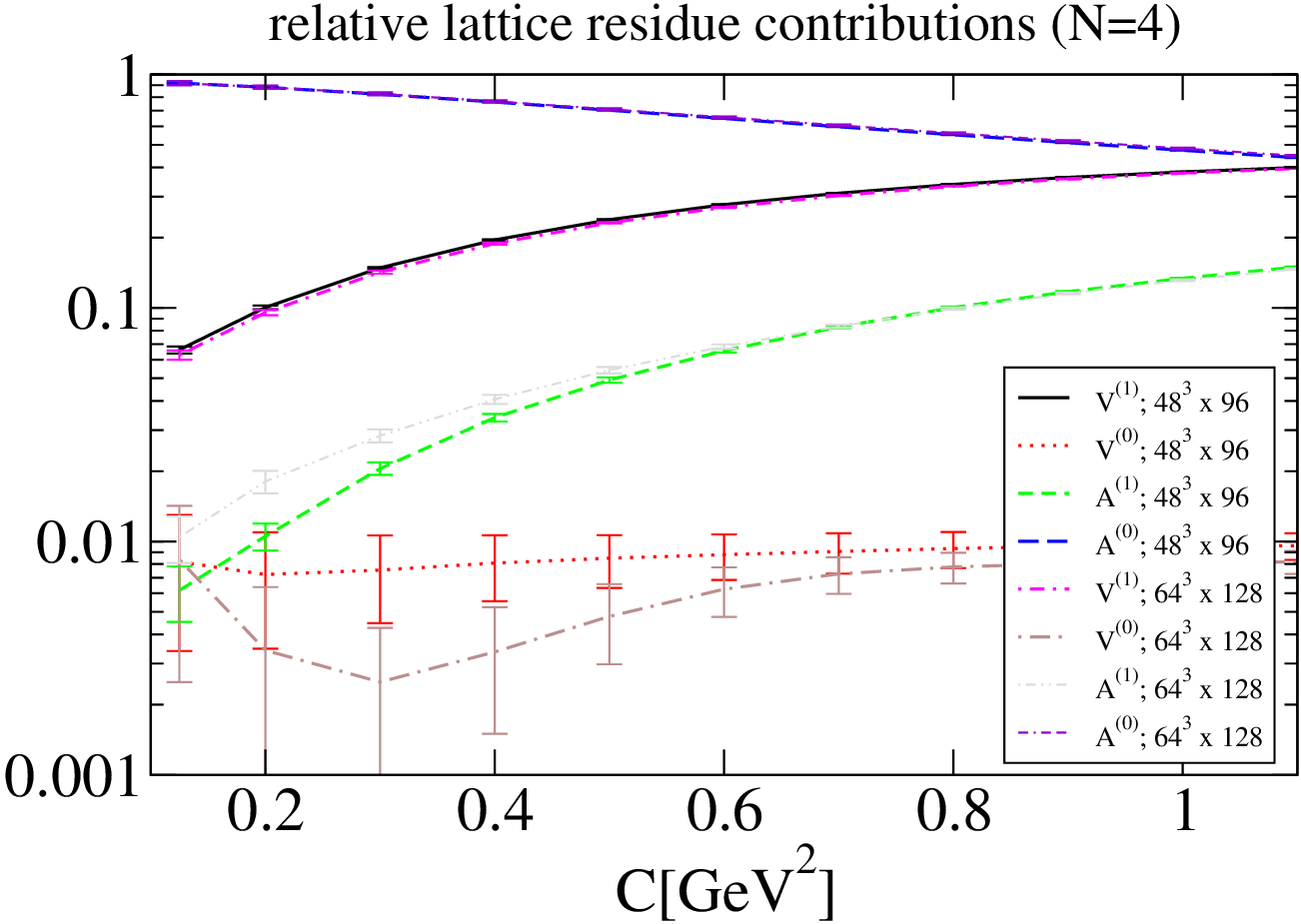} 
\includegraphics[scale=0.3,clip]{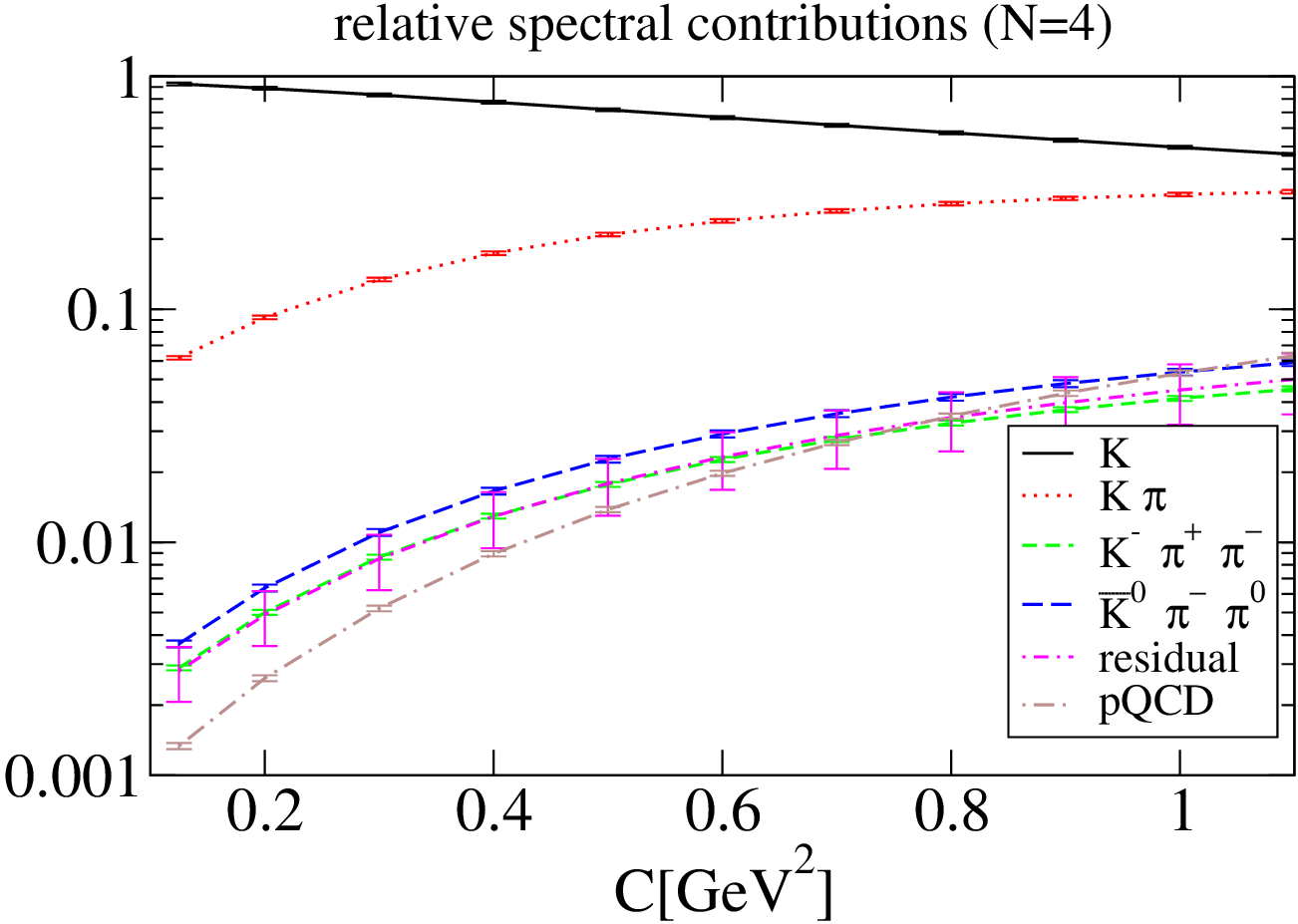} 
\caption{ 
Relative contributions of $J=0,1$, V and A channels to the sum of 
lattice residues (upper panel), and different (semi-) exclusive modes 
to the weighted $us$ spectral integrals (lower panel), as a function 
of $C$, for $N=4$.}
\label{fig:ratio}
\end{figure}

The upper panel of Fig.~\ref{fig:ratio} shows the relative sizes of 
the four $C$-dependent lattice contributions, $V^{(J)}$, $A^{(J)}$, 
for $N=4$. The lower panel, similarly, shows the relative sizes of 
different contributions to the weighted $us$ spectral integrals. 
$K\pi$ denotes the sum of $K^-\pi^0$ 
and $\bar{K}^0 \pi^-$ contributions, {\it pQCD} the contribution from 
$s > m_\tau^2$, evaluated using the 5-loop-truncated pQCD 
form~\cite{Baikov:2008jh,footnote}. Varying $C$ (and $N$) varies 
the level of suppression of the pQCD and higher-multiplicity 
contributions, the relative size of $K$ and $K\pi$ contributions, and 
hence the level of ``inclusiveness'' of the analysis. The stability of 
$\vert V_{us}\vert$ under such variations provides additional systematic 
cross-checks.

\section{Analysis and Results}
The $A^{(0)}$ channel produces the largest RHS contribution to 
Eq.~(\ref{eq:gd}). On the LHS, the $K$ pole dominates $\rho^{(0)}_{us;A}(s)$, 
with continuum contributions doubly chirally suppressed. Estimated LHS 
continuum $A^{(0)}$ contributions, obtained using sum-rule $K(1460)$ and 
$K(1830)$ decay constant results~\cite{kmsps}, are numerically 
negligible for the $\omega_N$ we employ. An ``exclusive'' $A^{(0)}$ 
analysis relating $\tilde{F}_{w_N}^{A^{(0)}}$ to the $K$-pole contribution, 
$\tilde{R}_{w_N}^K= \gamma_K \omega_N(m_K^2)$ is, therefore, possible,
with $\gamma_K = 2\vert V_{us}\vert^2 f_K^2$ obtained from either 
$K_{\mu2}$ or $\Gamma [\tau\rightarrow K\nu_\tau ]$.
Since the simulations underlying $\tilde{F}_{w_N}^{A^{(0)}}$ are isospin 
symmetric, we correct $\gamma_K$ for leading-order electromagnetic (EM) and 
strong isospin-breaking (IB) effects~\cite{Aoki:2016frl, Antonelli:2013usa}. 
With PDG $\tau$ lifetime~\cite{Olive:2016xmw} 
and HFLAV $\tau\rightarrow K\nu_\tau$ BF~\cite{HFLAV-Tau2017} input,
$\gamma_K[\tau_K]=0.0012061(167)_{exp}(13)_{IB}$ GeV$^2$.
$\gamma_K[\tau_K]$ is employed in our main, fully $\tau$-based analysis. 
The more precise result 
$\gamma_K[K_{\mu2}] = 0.0012347(29)_{exp}(22)_{IB}$~\cite{Olive:2016xmw} 
from $\Gamma [K_{\mu 2}]$ can also be used if SM dominance is assumed.
Exclusive analysis $\vert V_{us}\vert$ results are independent of $C$ for 
$C < 1$ GeV$^2$ (confirming tiny continuum $A^{(0)}$ contributions) and 
agree with the results, $\vert V_{us}\vert =0.2233(15)_{exp}(12)_{th}$ 
and $0.2260(3)_{exp}(12)_{th}$, obtained using 
$\vert V_{us}\vert = \sqrt{\gamma_K/(2f_K^2)}$, the isospin-symmetric 
lattice result $F_K\equiv \sqrt{2}f_K=0.15551(83)$ GeV~\cite{Blum:2014tka} 
and $\gamma_K =\gamma_K[\tau_K]$ and $\gamma_K[K_{\mu2}]$, respectively.
See \cite{footnote1} for further details.

For the fully inclusive analysis, statistical and systematic uncertainties 
are reduced by employing $2f_K^2 \omega_N(m_K^2)$, with measured $f_K$, for
the $K$ pole $A^{(0)}$ channel contribution. The residual, continuum 
$A^{(0)}$ contributions are compatible with zero within errors, as anticipated 
above. IB corrections, beyond those applied to $\gamma_K$, are numerically 
relevant only for $K \pi$. We account for (i) $\pi^0$-$\eta$ mixing, 
(ii) EM effects, and (iii) IB in the phase space factor, with $\pi^0$-$\eta$ 
mixing numerically dominant, evaluating these corrections, and their 
uncertainties, from the results presented in Ref.~\cite{Antonelli:2013usa}.
A $2\%$ uncertainty, estimated using results from a study of duality 
violations in the $SU(3)_F$-related flavor $ud$ channels~\cite{boito2014}, 
is assigned to pQCD contributions. Since our analysis is optimized 
for $\omega_N$ strongly suppressing higher-multiplicity and $s>m_\tau^2$ 
contributions, such an uncertainty plays a negligible role in
our final error.

Several systematic uncertainties enter the lattice computation. With an
assumed continuum extrapolation linear in $a^2$ but only two lattice 
spacings, $\mathcal{O}(a^4)$ discretization uncertainties must be estimated. 
For the $\omega_N$ we employ, the two ensembles yield $\tilde{F}_{\omega_N}$ 
differing by less than (typically significantly less than) $10 \%$, 
compatible with $\sim Ca^2$ or smaller $\mathcal{O}(a^2)$ errors. 
Anticipating a further $\sim Ca^2$ reduction of $\mathcal{O}(a^4)$ 
relative to $\mathcal{O}(a^2)$ corrections, we estimate residual 
$\mathcal{O}(a^4)$ continuum extrapolation uncertainties to be 
$\sim 0.1\, C a_f^2$, with $a_f^{-1}=2.36$ GeV~\cite{Blum:2014tka} 
the smaller of the two lattice spacings.
We also take into account the lattice scale setting uncertainty. The 
dominant FV effect is expected to come from $K\pi$ loop contributions 
in the $V^{(1)}$ channel, which we estimate using a lattice regularized 
version of finite-volume Chiral Perturbation Theory (ChPT). 
It is known, from Ref.~\cite{Aubin:2015rzx}, that one-loop ChPT for 
HVPs involving the light $u, d$ quarks yields a good semi-quantitative 
representation of observed FV effects~\cite{footnote2}; we thus expect it to also
work well for the flavor $us$ case considered here, where FV effects 
involving the heavier $s$ quark should be suppressed relative to those 
in the purely light $u, d$ quark sector. The result of our one-loop
ChPT estimate is a $1\%$ FV correction. We thus assign a 1\% FV
uncertainty to our $V^{(1)}$ channel contributions~\cite{footnote3}.
Regarding the impact of the slight $u,\, d,\, s$ 
sea-quark mass mistunings on the PQ results, the shift from 
slightly mis-tuned unitary to PQ shifted-valence-mass results for 
$\tilde{F}(\omega_N)$ corresponds to shifts in $\vert V_{us}\vert$ 
of $< 0.4\%$ for both ensembles. With masses and decay constants
typically much less sensitive to sea-quark mass shifts than 
to the same valence-quark mass shifts, we expect sea-mass PQ effects 
to be at the sub-$\sim 0.1\%$ level, and hence negligible on the
scale of the other errors in the analysis.

\begin{figure}[t]
\begin{center}
\includegraphics[scale=0.3,clip]{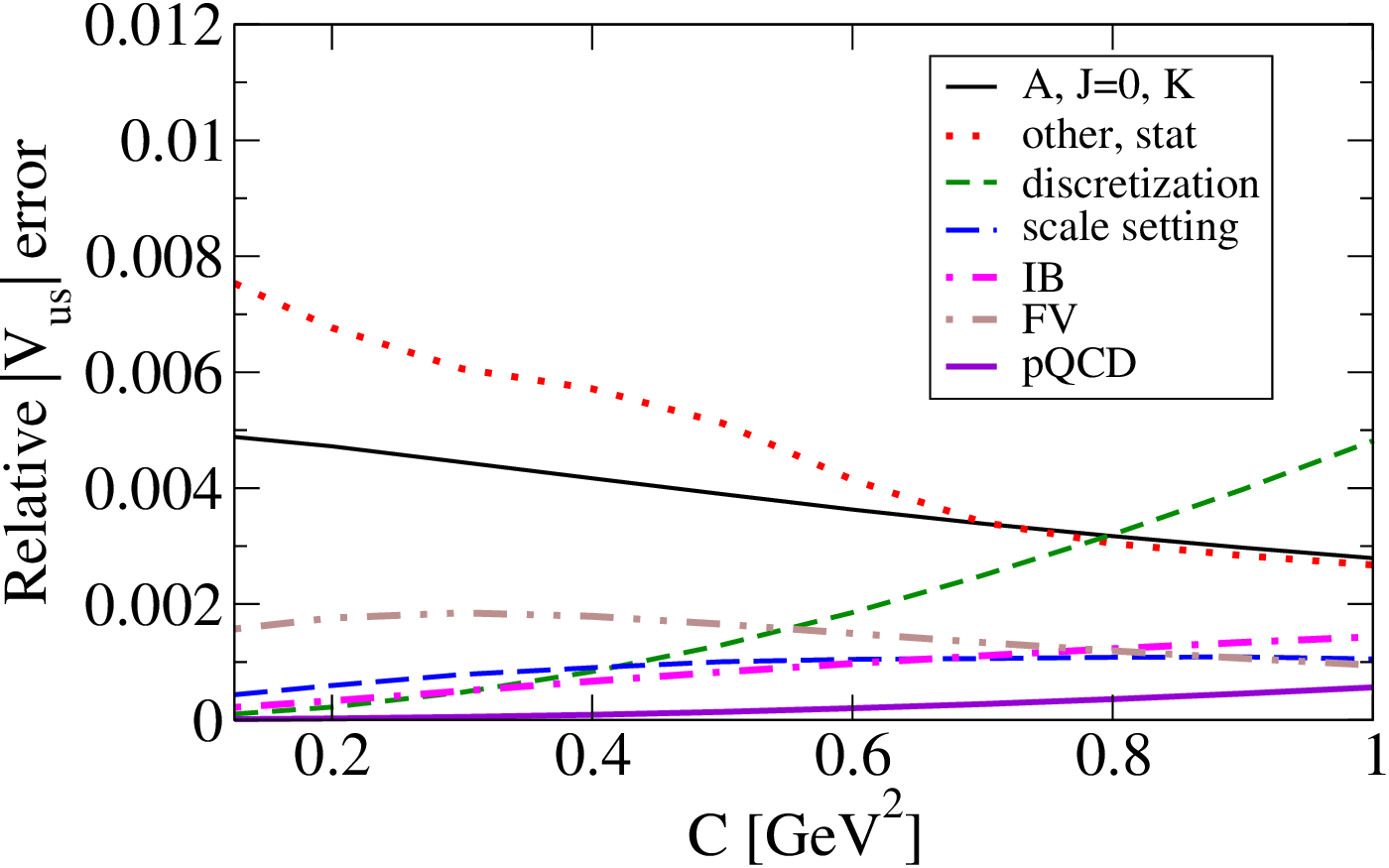}
\caption{ 
Lattice $\vert V_{us}\vert$ error contributions for $N=4$.}
\label{fig:rerr_residue}
\end{center} 
 \end{figure}

Fig.~\ref{fig:rerr_residue} shows the $C$ dependence of relative, 
non-data, inclusive analysis error contributions. $K$ labels the
$f_K$-induced $A^{(0)}$ uncertainty, {\it other} that induced by
the statistical error on the sum of $V^{(1)}$, $V^{(0)}$, $A^{(1)}$, 
and tiny continuum $A^{(0)}$ channel contributions. The statistical error 
dominates for low $C$, the discretization error for large $C$.

\begin{figure}[t]
\begin{center}
\includegraphics[scale=0.3,clip]{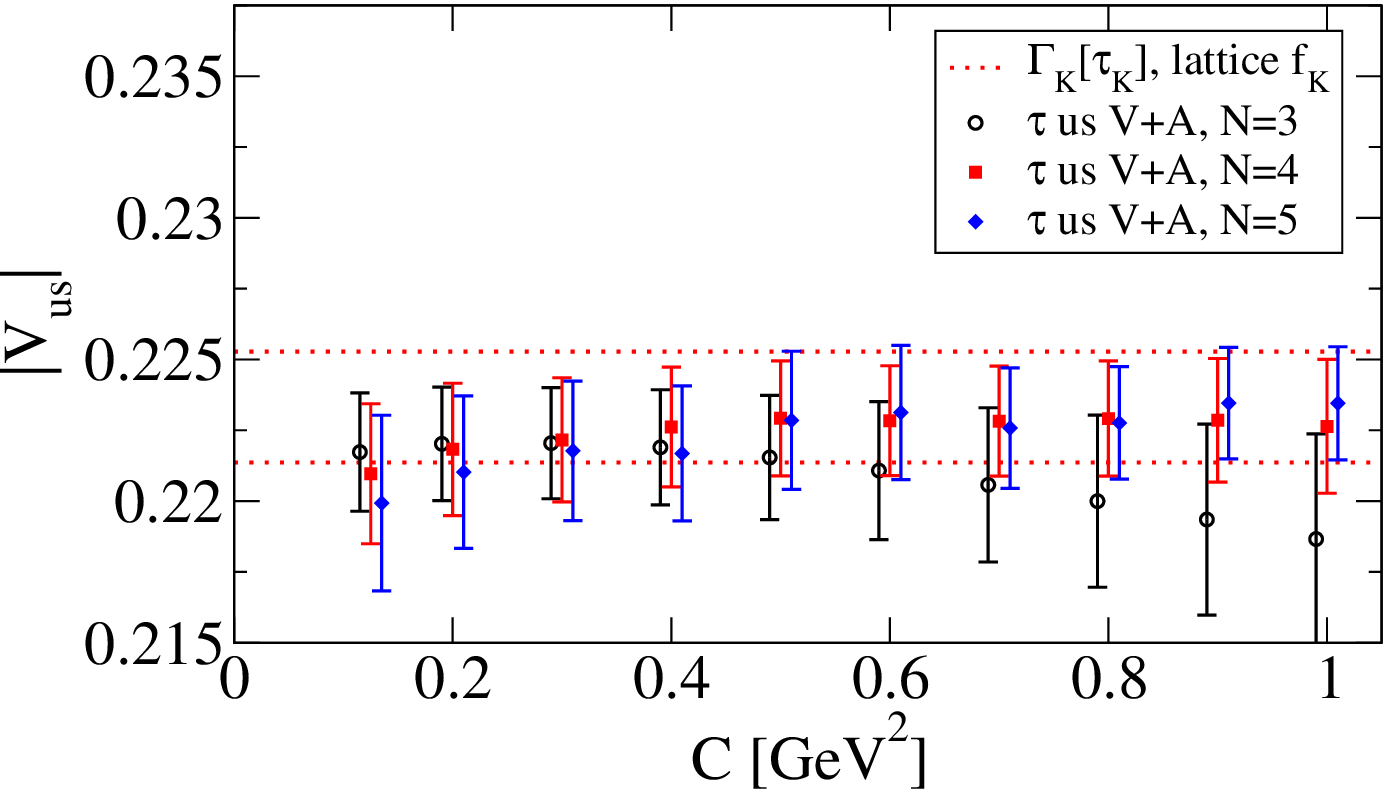} 
\caption{ 
$\vert V_{us}\vert$ vs. $C$ for $N=3,4,5$.
$N=3, 5$ results are shifted horizontally for presentational clarity and
statistical and systematic errors added in quadrature.
The determination using $\gamma [\tau_K]$ and lattice $f_K$
is shown for comparison.}
\label{fig:vus}
\end{center} 
 \end{figure}

Fig.~\ref{fig:vus} shows our $\vert V_{us}\vert$ results. These agree well 
for different $N$, and $C<1$ GeV$^2$. The slight trend toward lower central 
values for weights less strongly suppressing high-$s$ spectral contributions 
($N=3$ and higher $C$) suggests the residual mode distribution may be 
somewhat underestimated due to missing higher-multiplicity contributions. 
Such missing high-$s$ strength would also lower the $\vert V_{us}\vert$
obtained from FB FESR analyses. Table~\ref{tab:ratio} lists relative spectral 
integral contributions for selected $\omega_N$. Note the significantly 
larger ($6.8\%$ and $21\%$) residual mode and pQCD contributions for $N=3$ 
and $C=1$ GeV$^2$. Restricting $C$ to $<1$ GeV$^2$ keeps these from growing 
further and helps control higher-order discretization errors.
The error budget for various sample weight choices is summarized in Table~\ref{tab:error}.

\begin{figure}[t]
\begin{center}
\includegraphics[scale=0.3,clip]{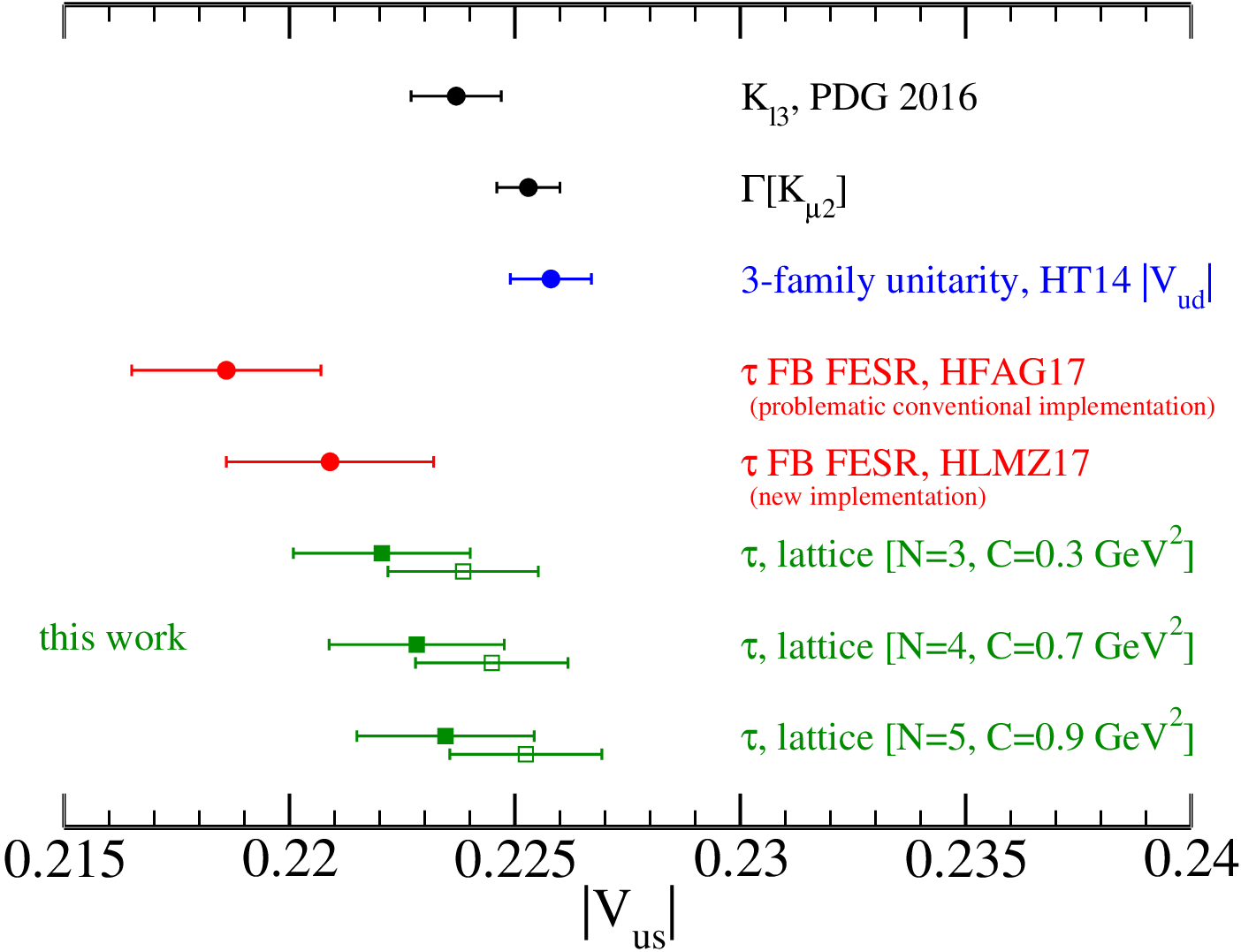} 
\caption{ 
Our $\vert V_{us}\vert$ determinations (inclusive, $\gamma_K[\tau_K]$-based 
exclusive (filled square) and $\gamma_K[K_{\mu2}]$-based exclusive 
(empty square)) c.f. results from other sources.}
\label{fig:summary}
\end{center} 
 \end{figure}

Our optimal inclusive determination is obtained for $N=4$, $C=0.7$ 
GeV$^2$, where residual mode and pQCD contributions are highly 
suppressed, and yields results
\begin{align}
\vert V_{us}\vert = 
\begin{cases}
0.2228(15)_{exp}(13)_{th}, & \rm{ for }\ \gamma_K[\tau_K] \\
0.2245(11)_{exp}(13)_{th}, & \rm{ for }\ \gamma_K[K_{\mu2}] \\
\end{cases}
\end{align}
consistent with determinations from $K$ physics and 3-family unitarity. 
Theoretical (lattice) errors are comparable to experimental ones and combined
errors improve on those of the corresponding inclusive FB FESR determinations. 
A comparison to the results of other determinations is given 
in Fig.~\ref{fig:summary}.

\begin{table}[htb]
  \centering
  \begin{tabular}{llrrrr}\hline\hline
contribution && \multicolumn{4}{c}{value[\%]} \\\hline
& [N, C[GeV$^2$]] & $[3, 0.3]$ & $[3, 1]$ & $[4, 0.7]$ & [5, 0.9] \\\hline
 $K$ && 65.5 & 30.9 & 61.7 & 66.9 \\
 $K\pi$ && 21.4 & 28.6 & 26.4 & 25.2 \\
 $K^-\pi^+\pi^-$ && 2.4 & 5.6 & 2.8 & 2.1 \\
 $\bar{K}^0\pi^-\pi^0$ && 3.1 & 7.3 & 3.6 & 2.7 \\
 residual && 2.7 & 6.8 & 2.9 & 2.1 \\
 pQCD && 4.9 & 20.8 & 2.7 & 1.1 \\\hline\hline
\end{tabular}
\caption{
Sample relative spectral integral contributions. }
  \label{tab:ratio}
\end{table}

\begin{table}[htb]
  \centering
  \begin{tabular}{llrrrr}\hline\hline
contribution && \multicolumn{4}{c}{relative error ($\%$)} 
\\\hline
& [N, C[GeV$^2$]] & $[3, 0.3]$ & \multicolumn{1}{r}{\ \ $[3, 1]$} & $[4, 0.7]$ & [5, 0.9]
\\\hline
theory & $f_K$ & 0.37 & 0.20 & 0.34 & 0.36 \\
& others, stat. & 0.41 & 0.19 & 0.34 & 0.41 \\
& discretization & 0.10 & 0.80 & 0.25 & 0.27 \\
& scale setting & 0.09 & 0.08 & 0.11 & 0.11 \\
& IB & 0.10 & 0.21 & 0.11 & 0.10 \\
& FV & 0.10 & 0.04 & 0.13 & 0.18 \\
& pQCD & 0.05 & 0.26 & 0.03 & 0.01 \\\hline 
& total & 0.59 & 0.91 & 0.58 & 0.65 \\
\hline\hline
experiment & $K$ &       0.48 &  0.27 &  0.44 &  0.47 \\ 
& $K\pi$ &       0.20 &  0.32 &  0.23 &  0.22 \\ 
& $K^-\pi^+\pi^-$ &      0.06 &  0.16 &  0.06 &  0.05 \\ 
& $\bar{K}^0\pi^-\pi^0$ &        0.03 &  0.09 &  0.03 &  0.03 \\ 
& residual &     0.41 &  1.35 &  0.41 &  0.28 \\\hline 
& total &        0.66 &  1.43 &  0.65 &  0.59 \\\hline\hline 
Combined & total &       0.88 &  1.70 &  0.87 &  0.88 \\\hline\hline 
  \end{tabular}
  \caption{
Error budget for the inclusive $\vert V_{us}\vert$ determination.}
  \label{tab:error}
\end{table}

\section{Conclusion and Discussion}
We have presented a novel method for determining $\vert V_{us}\vert$ using 
inclusive strange hadronic $\tau$ decay data.
Key advantages over the 
related FB FESR approach employing the same $us$ data are (i) the use of 
systematically improvable precision lattice data in place of the OPE, 
and (ii) the existence of weight functions that more effectively suppress 
spectral contributions from the larger-error, high-$s$ region 
without blowing up theory errors.
The results provide not only the most accurate inclusive
$\tau$ decay sum rule determination of $\vert V_{us}\vert$ but also
evidence that high-$s$ region systematic errors may be 
underestimated in the alternate FB FESR approach. 
The combined experimental uncertainty 
can be further reduced through improvements to the experimental 
$\tau\rightarrow K,\, K\pi$ BFs, while the largest of the current 
theoretical errors, that due to lattice statistics, is improvable by 
straightforward lattice computational effort. 
Such future improvements 
will help constrain the flavor dependence of any new physics contributions present in 
hadronic tau decays, contributions expected to be present at some level if 
the apparent violation of lepton flavor universality seen recently in 
semileptonic $b \to c$ decays involving $\tau$ persists.
\\
\acknowledgments 
We are grateful to RBC/UKQCD for fruitful discussion and support.
We thank the Benasque centre for hospitality at the Benasque workshop 
``High Precision QCD at low energy'', where this project started.
The research leading to these results has received funding from the 
European Research Council under the European Union's Seventh Framework 
Programme (FP7/2007-2013) / ERC Grant agreement 279757 and STFC ST/P000711/1. 
The authors gratefully acknowledge computing time granted through the 
STFC funded DiRAC facility (grants ST/K005790/1, ST/K005804/1, 
ST/K000411/1, ST/H008845/1). The software used includes the CPS QCD 
code (http://qcdoc.phys.columbia.edu/cps.html), 
supported in part by the USDOE SciDAC program; 
and the BAGEL (https://www2.ph.ed.ac.uk/\~{}paboyle/bagel/)
assembler kernel generator for high-performance optimised kernels 
and fermion solvers~\cite{Boyle:2009vp}.
This work is supported by resources provided by the Scientific 
Data and Computing Center (SDCC) at Brookhaven National Laboratory (BNL), 
a DOE Office of Science User Facility supported by the Office of Science 
of the U.S. Department of Energy.
The SDCC is a major component of the 
Computational Science Initiative (CSI) at BNL.
This is also supported in part
by JSPS KAKENHI Grant Numbers 17H02906 and 17K14309. 
HO is supported in part by RIKEN Special Postdoctoral Researcher program, 
Nara Women's University Intramural Grant for Project Research. 
RJH, RL and KM are supported by 
grants from the Natural Science and Engineering Research Council
of Canada. C.L. acknowledges support through a DOE Office of Science 
Early Career Award and by US DOE Contract DESC0012704(BNL).
A.P. is supported in part by UK STFC grants ST/L000458/1 and ST/P000630/1.

\clearpage

\setcounter{page}{1}
\renewcommand{\thepage}{Supplementary Information -- S\arabic{page}}
\setcounter{table}{0}
\renewcommand{\thetable}{S\,\Roman{table}}
\setcounter{equation}{0}
\renewcommand{\theequation}{S\,\arabic{equation}}

We show additional, supplementary plots and tables containing detailed 
information on the experimental distributions, expanding on the 
exclusive determination of $\vert V_{us}\vert$ using lattice
$A^{(0)}$ contributions and $f_K$ input, and providing the
ACLP~\cite{Antonelli:2013usa} $K\pi$ normalization analogues of the 
HFLAV~\cite{HFLAV-Tau2017} $K\pi$ normalization results shown in the main text.

\begin{figure}[htbp]
\begin{center}
\includegraphics[scale=0.3,clip]{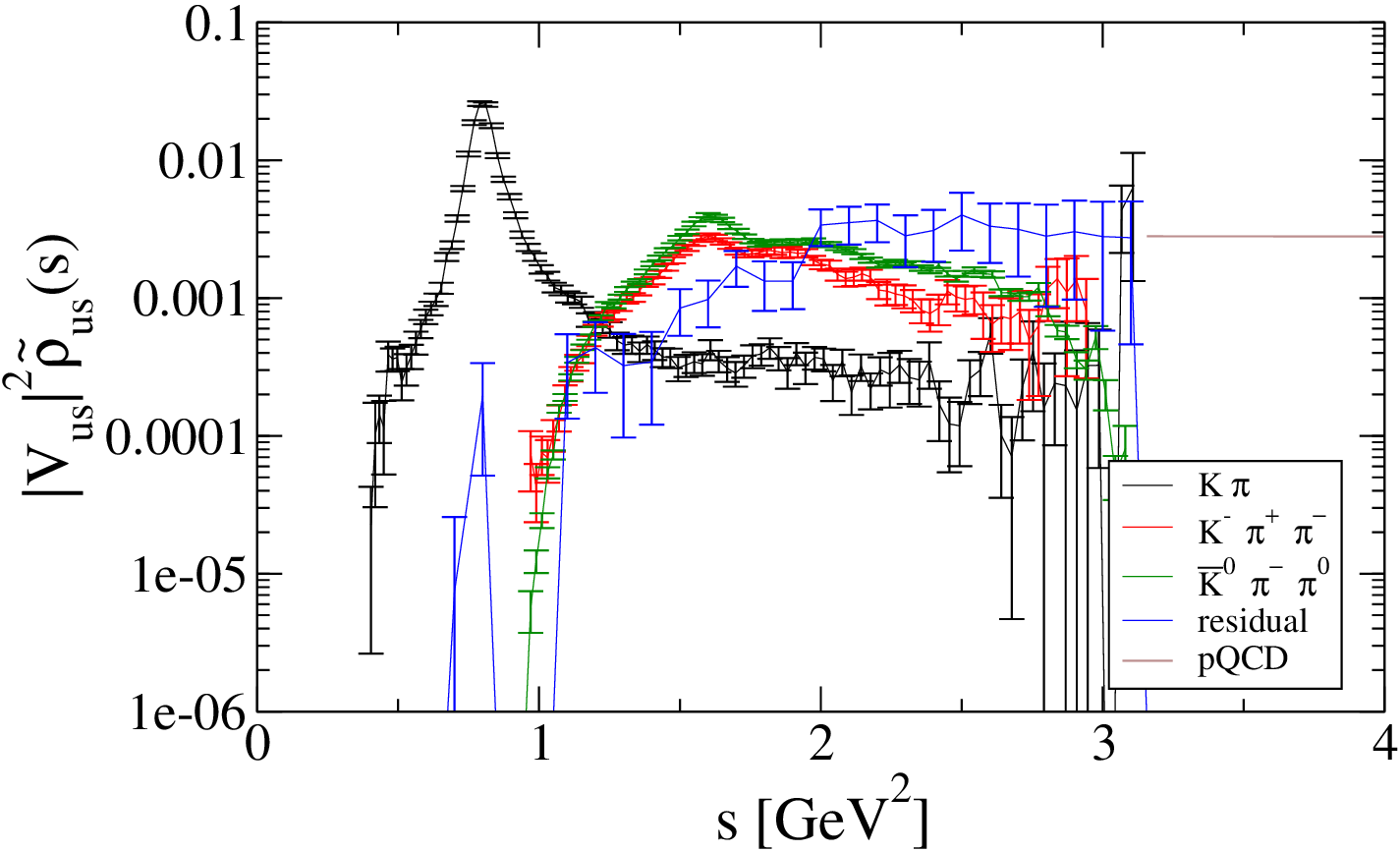} 
\includegraphics[scale=0.3,clip]{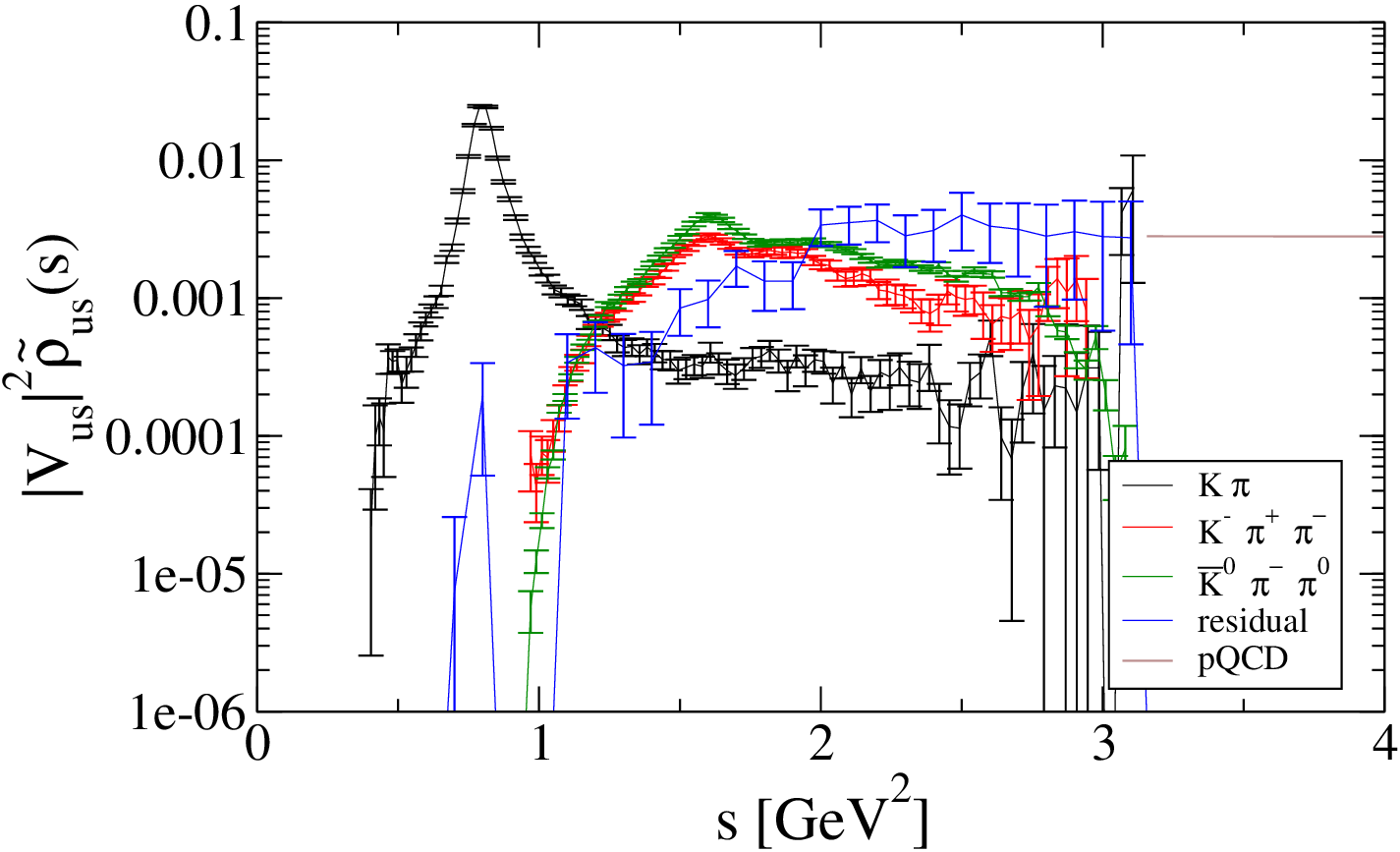} 
\caption{ 
Experimental exclusive-mode $\vert V_{us}\vert^2
\tilde{\rho}_{us}(s)$ contributions. Upper 
panel: ACLP $K\pi$ normalization, lower panel: HFLAV $K\pi$ normalization.}
\label{fig:exp}
\end{center} 
\end{figure}

\begin{figure}[htbp]
\begin{center}
\includegraphics[scale=0.3,clip]{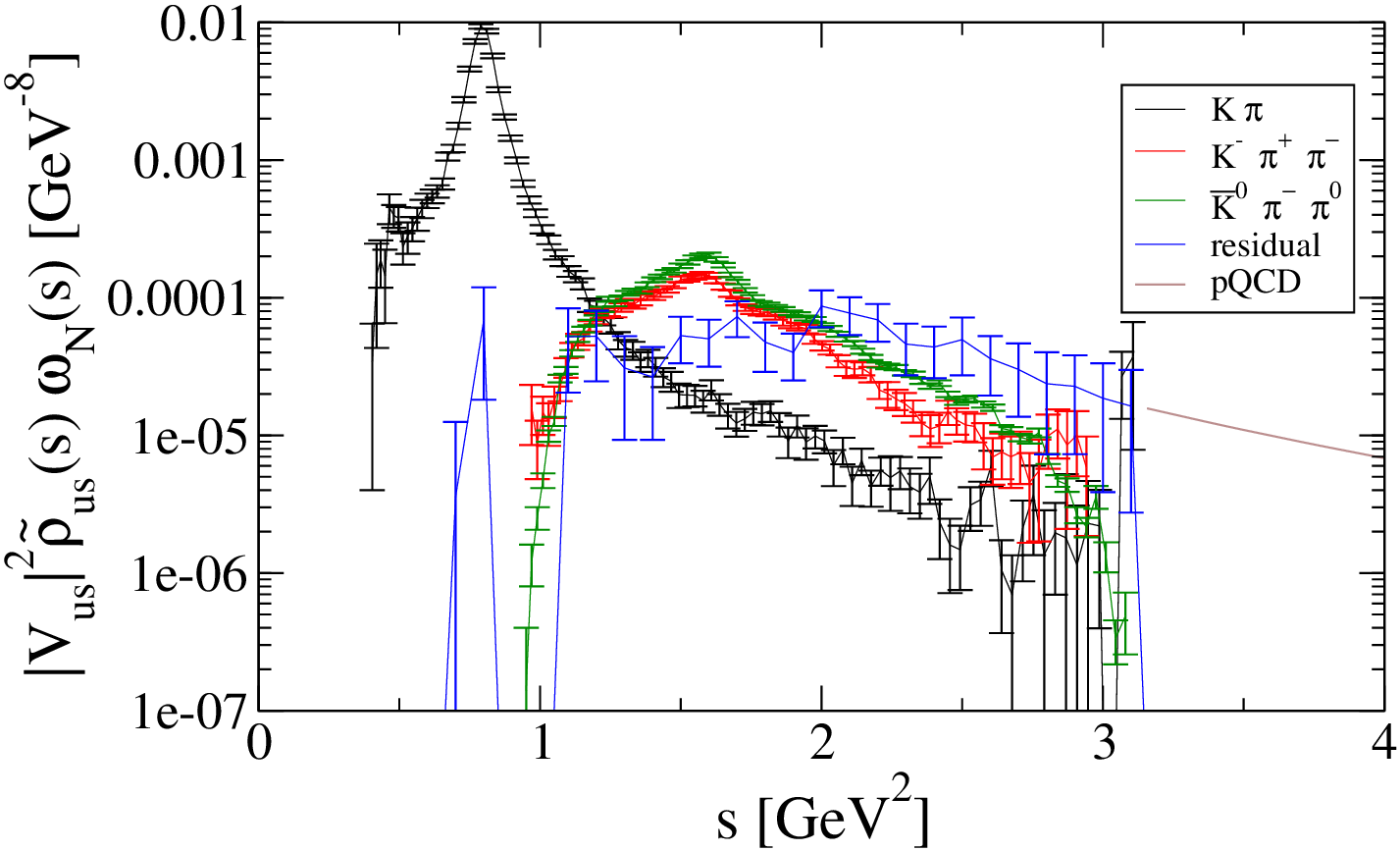} 
\includegraphics[scale=0.3,clip]{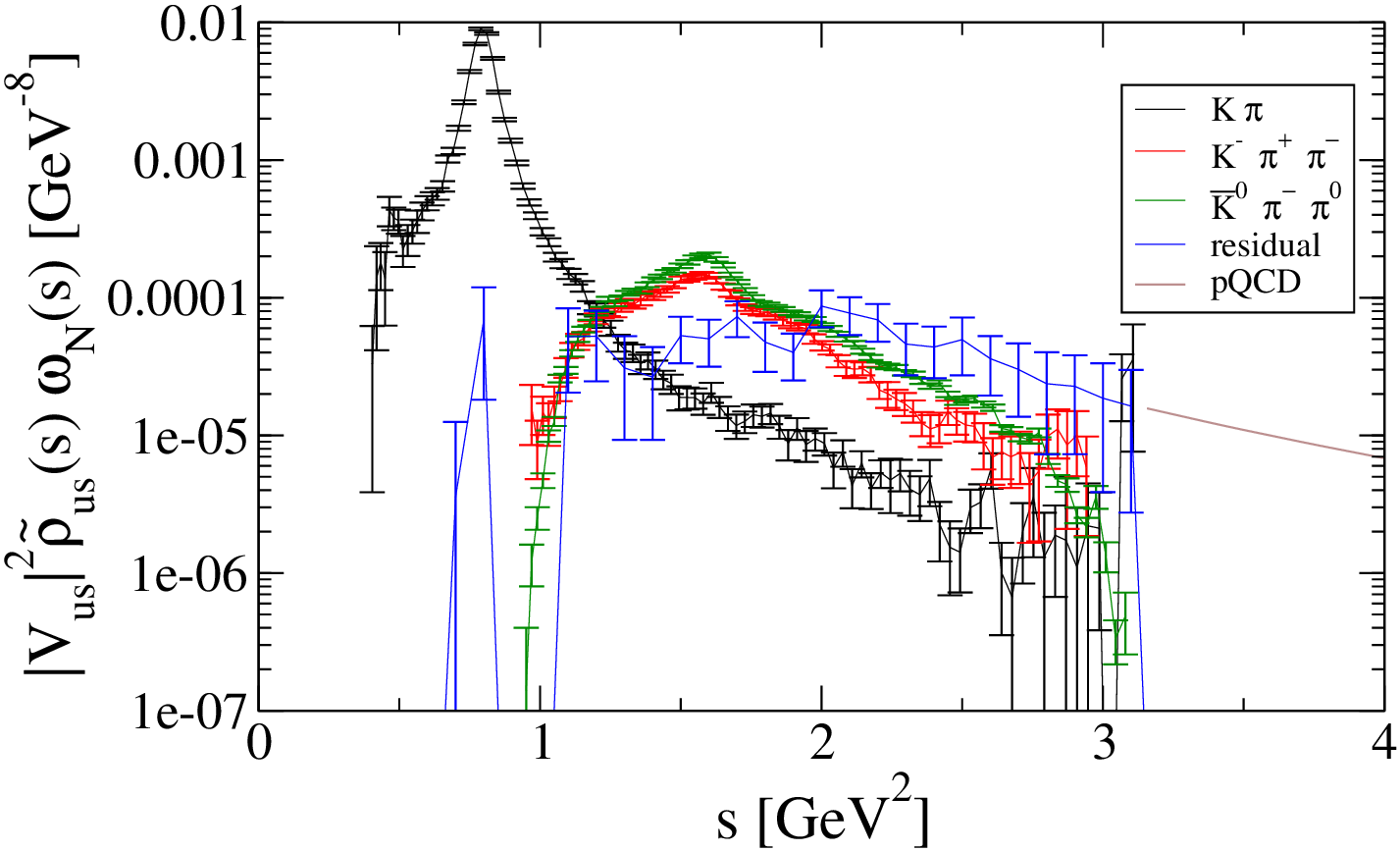} 
\caption{ 
Weighted experimental exclusive-mode ($\vert V_{us}\vert^2
\tilde{\rho}_{us}(s)\omega_N(s)$) 
contributions for the weight choice $N=4$ and $C=0.5$ GeV$^2$. Upper panel: 
ACLP $K\pi$ normalization, lower panel: HFLAV $K\pi$ normalization.}
\label{fig:exp-w}
\end{center} 
 \end{figure}

Fig.~\ref{fig:exp} shows the experimental exclusive mode distributions
for both the HFLAV and ACLP $K\pi$ normalization choices.
Fig.~\ref{fig:exp-w} shows an example of the weighted versions thereof,
for the $N=4$ and $C=0.5$ GeV$^2$ weight choice. The figures include, 
for reference, the pQCD contribution, evaluated using the 5-loop-truncated 
$D=0$ OPE form~\cite{Baikov:2008jh} and, for illustrative purposes, the 
sample value $\vert V_{us}\vert =0.2253$.  

The experimental results for $\vert V_{us}\vert^2 \tilde{\rho}_{us}(s)$ 
are obtained as a sum over exclusive and semi-exclusive mode contributions. 
B-factory unit-normalized, ${\frac{1}{N}}{\frac{dN}{ds}}$, distributions 
are used for the $K\pi$~\cite{Epifanov:2007rf}, 
$K^-\pi^+\pi^-$~\cite{Nugent:2013ij}, $\bar{K}^0 \pi^-\pi^0$~\cite{Ryu:2013lca}
and $K^-K^+K^-$~\cite{Nugent:2013ij} modes. The corresponding exclusive-mode 
$dR_{us;V+A}/ds$ contributions are obtained by multiplying the unit-normalized 
$\tau\rightarrow X\nu_\tau$ distributions by $B_X/B_e$, with $B_e$ the 
$\tau\rightarrow e\bar{\nu}_e\nu_\tau$ branching fraction (BF) and $B_X$ that 
of $\tau\rightarrow X\nu_\tau$. In the case of the two $K\pi$ modes, 
the preliminary BaBar results for the $K^-\pi^0$ distribution, reported in 
Ref.~\cite{Aubert:2007mh}, were never finalized. With the unit-normalized 
$K^-\pi^0$ and $K_S\pi^-$ distributions identical in the isospin limit, 
this problem can be dealt with, up to small isospin-breaking 
corrections, by using the unit-normalized Belle $K_S \pi^-$ 
distribution~\cite{Epifanov:2007rf} for both $K\pi$ modes. This leads to
\begin{equation}
{\frac{dR_{\bar{K}\pi}(s)}{ds}}=
\left( {\frac{B_{K^-\pi^0}+B_{\bar{K}^0\pi^-}}{B_e}}\right)
\left[\frac{1}{N} \frac{dN}{ds}\right]_{K_s\pi^-}\, .
\nonumber\end{equation}
As noted in the main text, two choices exist for the 2-mode $K\pi$ BF sum, 
that given by HFLAV~\cite{HFLAV-Tau2017} and that obtained from the extended 
analyis of ACLP~\cite{Antonelli:2013usa}, in which additional dispersive 
constraints were imposed on the timelike $K\pi$ form factors (at the
cost of somewhat larger final errors on the two $K\pi$ BFs). HFLAV 
BFs~\cite{HFLAV-Tau2017} are used for all other modes. For $K^-\pi^+\pi^-$,
we employ the full covariances of the unit-normalized distribution,
provided by BaBar~\cite{Nugent:2013ij}. The errors on the corresponding 
weighted spectral integrals turn out to be dominated by the BF uncertainty. 
For example, for the weight choice $N=4$, $C=0.5$ GeV$^2$, $2.29\%$ of the 
total $2.42\%$ relative error comes from the uncertainty in the 2-mode 
$K\pi$ BF sum. It is also worth noting that correlations play a relatively 
minor role in determining the small remaining contribution to the total 
error; including the correlations in the unit-normalized distribution 
results shifts this (already small) contribution by only $25\%$. For
the $\bar{K}^0\pi^-\pi^0$ mode, at present only the errors on the
unit-normalized distribution are known. Ignoring the (presently unknown)
correlations, one finds, for the weight choice $N=4$, $C=0.5$ GeV$^2$,
$3.38\%$ of the total $3.41\%$ relative error generated by the HFLAV
BF uncertainty. In view of the minor role played by correlations in 
determining the error on the exclusive $K^-\pi^+\pi^-$ contribution, 
we conclude that our lack of knowledge of the correlations in the 
unit-normalized $\bar{K}^0\pi^-\pi^0$ distribution will have a negligible 
impact on final error on contributions for this exclusive mode.
The residual mode $dR_{us;V+A}/ds$ distribution is obtained by summing 
1999 ALEPH results~\cite{Barate:1999hj} for those modes ($K^-\pi^0\pi^0$, 
$K\, 3\pi$, $K\, \eta$, $K\, 4\pi$ and $K\, 5\pi$) included in the ALEPH 
distribution but not remeasured by either Belle or BaBar. Each such 
contribution is first rescaled by the ratio of the current to the 1999 
ALEPH BF for the mode (or sum of modes), and the rescaled contributions 
then re-summed to produce the updated version of the residual mode 
distribution. 

Exclusive and semi-exclusive mode contributions to the product 
$\vert V_{us}\vert^2\tilde{\rho}_{us}(s)$ are obtained from the 
corresponding contributions to $dR_{us;V+A}/ds$ by division by 
$12\pi^2 S_{EW}\left( 1-{\frac{s}{m_\tau^2}}\right)^2/m_\tau^2$.

\begin{figure}[htbp]
\begin{center}
\includegraphics[scale=0.3,clip]{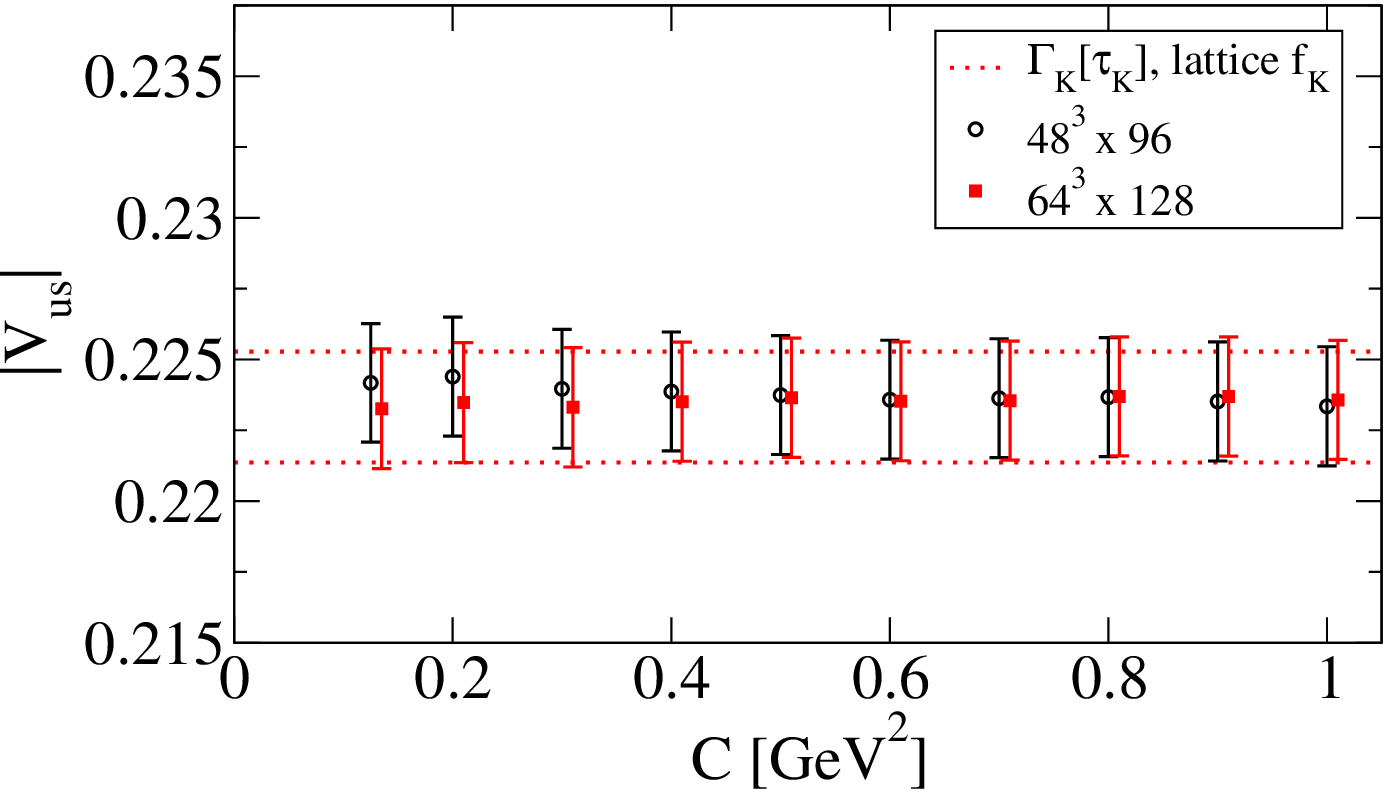} 
\caption{ 
$\vert V_{us}^{A_0}\vert$ as a function of $C$ for $N=4$, using 
experimental $\gamma_K[\tau_K]$ input. The error is statistical only.
The determination obtained using $\gamma[\tau_K]$, but also inputting
lattice $f_K$, is shown for comparison.}
\label{fig:vus_A0}
\end{center} 
 \end{figure}

Fig.~\ref{fig:vus_A0} shows the result of the exclusive $A^{(0)}$ 
determination of $\vert V_{us}^{A_0}\vert=\sqrt{\tilde{R}_{\omega_N}^K/
{\tilde{F}_{\omega_N}^{A^{(0)}}}}$ for $N=4$ as an example. 

\begin{figure}[t]
\includegraphics[scale=0.3,clip]{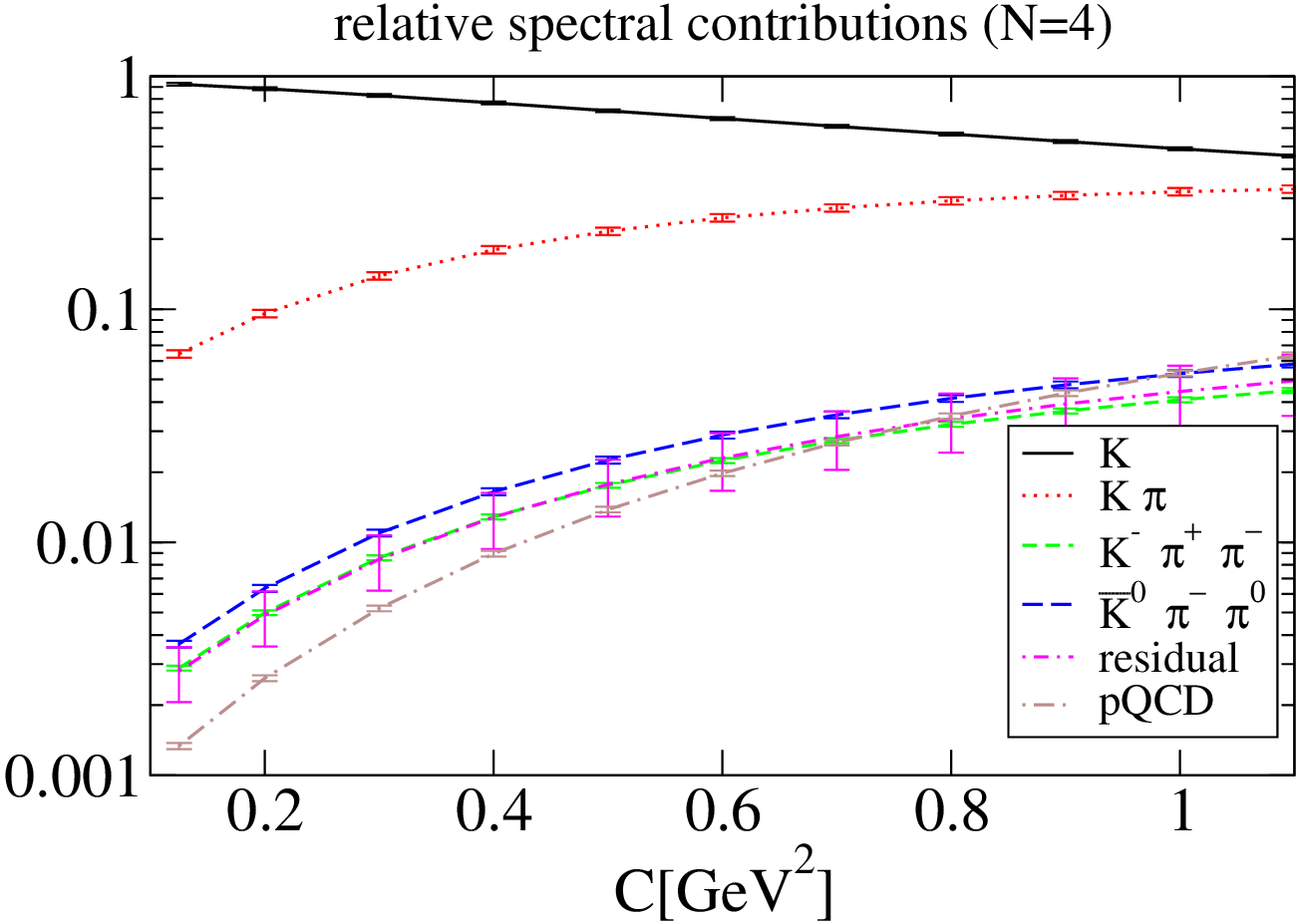}
\caption{
Relative contributions of different (semi-) exclusive modes
to the weighted $us$ spectral integrals, as a function of $C$, for $N=4$
and the ACLP $K\pi$ normalization.}
\label{fig:ratio_aclp}
\end{figure}

Fig.~\ref{fig:ratio_aclp} shows the relative contributions of the different
exclusive and semi-exclusive modes to the $N=4$ weighted $us$ spectral 
integrals, as a function of $C$, for the ACLP $K\pi$ normalization
choice (the corresponding HFLAV $K\pi$ normalization results are
shown in the bottom panel of Fig.~2 in the main text).

\begin{figure}[t]
\begin{center}
\includegraphics[scale=0.3,clip]{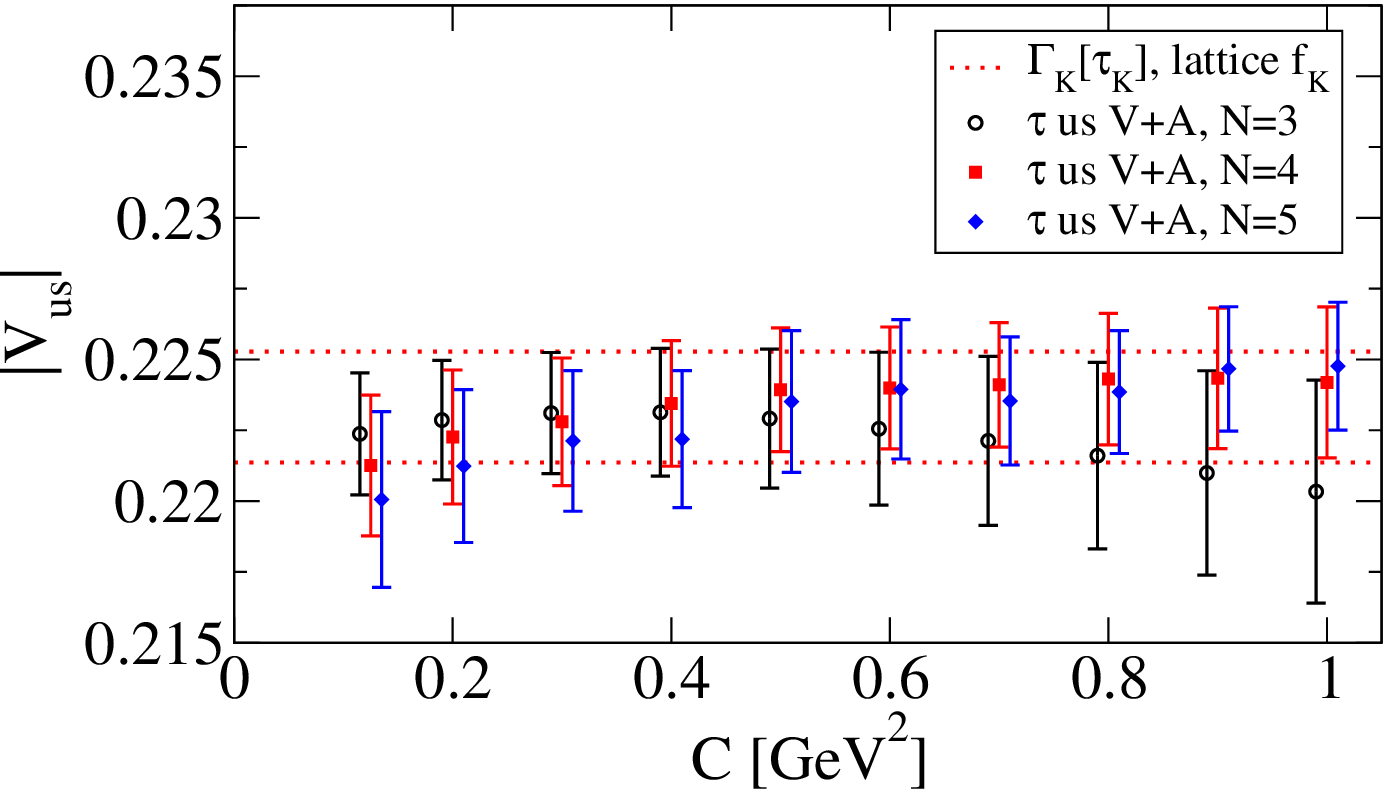}
\caption{
$\vert V_{us}\vert$ vs. $C$ for $N=3,4,5$, for the ACLP $K\pi$ normalization.
$N=3, 5$ results are shifted horizontally for presentational clarity and
statistical and systematic errors added in quadrature.
The determination using $\gamma [\tau_K]$ and lattice $f_K$
is shown for comparison.}
\label{fig:vus_hfag}
\end{center}
 \end{figure}

Fig.~\ref{fig:vus_hfag} similarly shows the results of the inclusive 
determination of $\vert V_{us}\vert$ as a function of $C$ for $N=3,4,5$, 
and the ACLP $K\pi$ normalization (the corresponding HFLAV $K\pi$ 
normalization results are shown in Fig.~4 of the main text).

Table~\ref{tab:ratio_aclp} shows the relative contributions of the 
various exclusive and semi-exclusive modes to the weighted $us$ 
spectral integrals for some sample weight choices, for the
ACLP $K\pi$ normalization case. (The HFLAV $K\pi$ normalization 
results for the same weight choices are shown in Table I of the main 
text.)
\begin{table}[htb]
  \centering
  \begin{tabular}{llrrrr}\hline\hline
contribution && \multicolumn{4}{c}{value[\%]} \\\hline
& [N, C[GeV$^2$]] & $[3, 0.3]$ & $[3, 1]$ & $[4, 0.7]$ & [5, 0.9] \\\hline
 $K$ && 64.9 & 30.4 & 61.0 & 66.2 \\
 $K\pi$ && 22.1 & 29.3 & 27.2 & 26.0 \\
 $K^-\pi^+\pi^-$ && 2.4 & 5.5 & 2.7 & 2.1 \\
 $\bar{K}^0\pi^-\pi^0$ && 3.1 & 7.2 & 3.5 & 2.7 \\
 residual && 2.6 & 6.7 & 2.8 & 2.0 \\
 pQCD && 4.9 & 20.8 & 2.7 & 1.1 \\\hline\hline
\end{tabular}
\caption{
Sample relative spectral integral contributions, ACLP $K\pi$ normalization.}
  \label{tab:ratio_aclp}
\end{table}

Table~\ref{tab:error_aclp} shows the experimental error budget obtained
using the ACLP $K\pi$ normalization for various sample weight choices.
(Results for the same weight choices and the HFLAV $K\pi$ normalization
are shown in the bottom half of Table II in the main text.) The last
line of the table gives the quadrature sums of the total experimental 
error and corresponding total theory error, the latter shown in the 
upper half of Table II in the main text.

\begin{table}[htb]
  \centering
  \begin{tabular}{llrrrr}\hline\hline
contribution && \multicolumn{4}{c}{relative error ($\%$)}
\\\hline
& [N, C[GeV$^2$]] & $[3, 0.3]$ & \multicolumn{1}{r}{\ \ $[3, 1]$} & $[4, 0.7]$
\
& [5, 0.9]
\\\hline
experiment & $K$ & 0.47 & 0.27 & 0.43 & 0.46 \\
& $K\pi$ & 0.43 & 0.69 & 0.51 & 0.48 \\
& $K^-\pi^+\pi^-$ & 0.06 & 0.15 & 0.06 & 0.05 \\
& $\bar{K}^0\pi^-\pi^0$ & 0.03 & 0.09 & 0.03 & 0.03 \\
& residual & 0.40 & 1.33 & 0.41 & 0.27  \\\hline
& total & 0.76 & 1.53 & 0.79 & 0.72 \\\hline\hline
Combined & total & 0.96 & 1.79 & 0.98 & 0.98 \\\hline
  \end{tabular}
  \caption{
Error budget for experimental contributions to the inclusive 
$\vert V_{us}\vert$ determination with ACLP $K\pi$ normalization.}
  \label{tab:error_aclp}
\end{table}

\begin{figure}[t]
\begin{center}
\includegraphics[scale=0.3,clip]{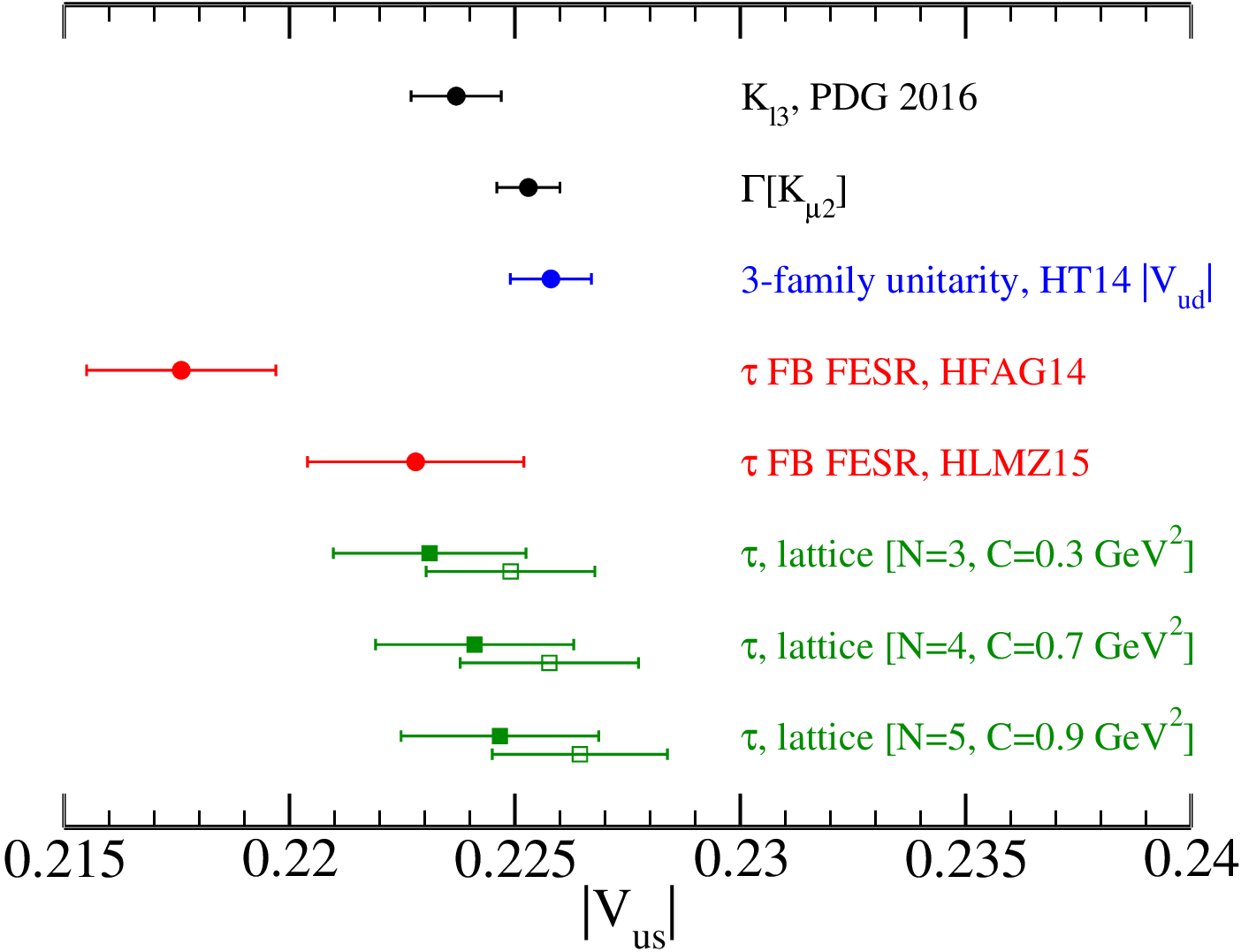}
\caption{
Our $\vert V_{us}\vert$ determinations (inclusive, $\gamma_K[\tau_K]$-based
exclusive (filled square) and $\gamma_K[K_{\mu2}]$-based exclusive
(empty square)) c.f. other determination results for the ACLP $K\pi$ 
normalization case.}
\label{fig:summary_aclp}
\end{center}
 \end{figure}

Finally, in Fig.~\ref{fig:summary_aclp}, we show the comparison of our 
ACLP-$K\pi$-normalization-based results for $\vert V_{us}\vert$ to those 
obtained by other methods, analogous to the HFLAV-$K\pi$-normalization-based 
comparison shown in Fig.~ 5 of the main text.

\clearpage


\begin{thebibliography}{99}
\bibitem{footnote}
$D=2$ OPE contributions have been checked and found to be numerically 
tiny compared to $D=0$ contributions.
\end{thebibliography}
\end{document}